\documentstyle[aps,pra,subfigure,epsfig,amsmath,amssymb]{revtex}
\begin{document}
\draft

\newcommand {\be}{\begin{equation}}
\newcommand {\ee}{\end{equation}}
\newcommand {\bea}{\begin{eqnarray}}
\newcommand {\eea}{\end{eqnarray}}
\newcommand {\refeq}[1] {(\ref{#1})}
\newcommand {\vett}[1] {{\mathbf{#1}}}
\newcommand {\rvec}{{\bf r}}

\title{Scissors mode in  a superfluid Fermi gas}
\author{Anna Minguzzi and Mario P. Tosi}
\address{Istituto Nazionale di Fisica della Materia and
Classe di Scienze, Scuola Normale Superiore, Piazza dei Cavalieri 7,
56126 Pisa, Italy}

\maketitle                
\begin{abstract}
We evaluate the frequencies of scissors modes for density and
concentration fluctuations in a vapour of fermionic atoms placed
in two hyperfine levels inside a spherical harmonic trap. Both the
superfluid and the normal state are considered, with  inclusion of
the interactions at the random-phase level. Two main results are
obtained: (i) the transition to the superfluid state is signalled by
the disappearance of soft transverse modes of the normal fluid in the
collisionless regime and (ii) the eigenfrequency of the density
fluctuations in the superfluid coincides with that of the normal fluid
in the collisional regime. The latter property is related to the opening of
the gap in the single-pair spectrum.
\end{abstract}


\section{Introduction}
\label{zero}
The possibility of obtaining a novel superfluid system by pairing of
ultra-cold fermionic alkali atoms seems to become closer to
experimental realization \cite{fermi_degen,gabriele}. Various mechanisms of
pairing have already  been proposed
\cite{stoof_prl,stoof_marianne,baranov_bcs}, 
the simplest one 
being an $s$-wave pairing between atoms belonging to two different
hyperfine levels of a magnetic trap. 

Revealing  the
superfluid transition in a Fermi gas is expected to be more
difficult than for bosonic condensation, since it cannot be inferred
from the observation of the 
density profile of the cloud \cite{stoof_marianne}. One has therefore
to look for the transition in
dynamical or kinematical properties of the fluid. We suggest in this
work a
possible method  for the detection  of the superfluid character of the
transition. The same test has already been successfully employed in
the case of a Bose-Einstein condensate \cite{david,onofrio}.

The property of superfluidity in a vapour is related by definition to
the character of its  dynamical response to a long-wavelength transverse
probe \cite{forster}, which will excite only the non-superfluid component.
In an inhomogeneous finite system, such as an
atomic vapour in magnetic or optical confinement or an atomic nucleus, a
test of the superfluidity can be obtained from the 
excitation of a small-angle oscillation in a plane
where the confinement is slightly anisotropic. For a
Bose-condensed gas this {\it scissors mode}
is predicted to show only one frequency component in the superfluid
state, 
 while it has two frequency components in the normal fluid due
to the additional contribution of  transverse excitations.  The
study of the scissors mode in Bose-Einstein condensates of alkali
vapours  has been suggested by 
Gu{\'e}ry-Odelin and Stringari \cite{david} and an experimental
realization of their ideas has already been given by Marag{\`o} {\it
et al.} \cite{onofrio}. 

The low-energy collective excitations of a trapped  Fermi
gas in the superfluid state  have been investigated by Baranov and Petrov \cite{baranov} in the
dilute (non-interacting) limit. Their theory is formulated 
 in terms of the fluctuations in
the phase of the order parameter, but they  show that these
modes also manifest themselves  as density fluctuations in
the sample. Their equation of motion for the
density fluctuations of a superfluid Fermi gas coincides with that obtained
by
Amoruso {\it et al.} \cite{ilaria} for a
non-interacting  normal Fermi fluid in the collisional regime, within a
local-density approximation (LDA). 

In this paper we derive in Sect.~\ref{uno} an equation of motion for
the density  
fluctuations of a superfluid Fermi gas in an improved LDA, which
includes also the effects of the 
interactions within a random-phase approximation (RPA) as already
proposed in early work by Anderson
\cite{anderson_rpa}. We make use of 
this equation in Sect.~\ref{due} to give theoretical
predictions for the excitation frequency of the scissors mode in the 
superfluid Fermi gas. This result is compared in Sect.~\ref{tre}
 with the excitation frequencies of the  scissors modes of
 a normal Fermi fluid in the degenerate regime and in the
presence of interactions. Finally, Sect.~\ref{quattro} presents a summary and
some concluding remarks.

\section{Coherent dynamics of a superfluid Fermi gas}
\label{uno}
The low-energy excitation spectrum of a neutral Fermi gas
with weak attractive forces at very low temperatures is characterized
in the superfluid phase by a longitudinal collective mode with a
linear dispersion relation at the velocity of ordinary (first)
sound. This mode is not allowed to propagate in the normal
phase as it is embedded in the particle-hole
continuum of excitations (just as it is the case for the zero-sound
mode in a fluid with 
attractive interactions). However, in  the
superfluid phase the opening of the gap $\Delta$ shifts the
energy threshold for the quasiparticle-quasihole continuum to
$2\Delta$, leaving a window in energy where  first sound is
stable.This mode is known as the 
Bogolubov-Anderson sound \cite{anderson_rpa,bogolubov_sound} and can be viewed
as a coherent oscillation of the order parameter for the condensed
phase; in fact, its presence is required by the Goldstone theorem to ensure
gauge invariance. The simplest theory which includes the
Bogolubov-Anderson mode in the density fluctuation spectrum is the random-phase
approximation for a superconductor \cite{anderson_rpa,nambu_rpa,griffin_rpa}. 

In a harmonically trapped superfluid  the Bogolubov-Anderson
sound becomes quantized due to the confinement. 
In the dilute limit the equation of motion for the density fluctuations
has been obtained by Baranov and Petrov \cite{baranov} using a semiclassical 
solution of the Bogolubov - de Gennes equations.
As already noted in Sect.~\ref{zero}, the resulting spectrum of
excitations coincides with that obtained  \cite{ilaria}  for an
harmonically trapped normal Fermi gas in the
collisional regime, 
when a local-density approximation for the kinetic stress tensor is
employed.

The coincidence of the two spectra can be explained --
within the LDA  -- by making use of the result for
the homogeneous fluid that the compressibility is unaffected by the
phase transition \cite{leggett}: a superfluid and a normal fluid at the
same density and interaction strength show the same value for the
compressibility. Indeed this property is not restricted
to the non-interacting Fermi gas: the coincidence of
the sound velocity in a superfluid and in a normal fluid in the
collisional regime has also been demonstrated by Larkin and Migdal
\cite{migdal} for generic interaction strengths within the theory of
superfluid Fermi liquids. For the specific case of a weakly
interacting Fermi gas the result for the sound
velocity $c$ in the normal fluid in the collisional regime, given by
$c^2=(v_f^2/3)(1+2 k_f a_{\uparrow\downarrow}/\pi)$ coincides with the
velocity of the Bogolubov-Anderson mode, as obtained from the
RPA \cite{paris-futur}. Here, $v_f$ and $k_f$ are the Fermi velocity
and wavenumber, while $a_{\uparrow\downarrow}$ is the $s$-wave
scattering length  between fermions in different hyperfine states. 

Another characteristic of the spectrum of a homogeneous Fermi
superfluid is that all  collective modes other than the density
fluctuations, such as spin-density and transverse current-density
modes,  are
suppressed in the zero temperature limit, since they are related to
the motion of the normal component of the fluid \cite{leggett2,martin_rpa}.   

These observations allow us to conclude that  the
equations of motion 
for the total density fluctuations in a symmetric 
two-component normal 
Fermi gas  
in the collisional regime
\cite{ilaria2} describe as well the collective excitations of a weakly
interacting superfluid at $T=0$
within a local-density approximation. Specifically, we take as the basic
equations for the dynamics of a superfluid Fermi gas in the linear
regime the following:
\begin{equation}
\partial_t n(\vett r,t)=-\nabla \cdot (n_{eq} (\vett r)
\vett v (\vett r,t)) \label{cont}
\end{equation}
and
\begin{equation}
\partial_t \vett v (\vett r,t)=-\nabla \left[\dfrac{1}{3}A
(n_{eq}(\vett r)/2)^{-1/3}n(\vett r,t)+\frac{1}{2} g n(\vett
r,t)\right]\;.\label{eul} 
\end{equation}
Here $A=\hbar^2 (6 \pi^2)^{2/3}/2m$, $g=4 \pi \hbar^2
a_{\uparrow\downarrow}/m$, and $n(\vett r,t)$ and $\vett
v(\vett r,t)$ are the total density fluctuation and velocity field for
the gas,  $n_{eq}(\vett r)$ being the equilibrium density profile.
Interactions have been included in the
theory at mean field (Hartree) level and exchange is not allowed since
$s$-wave interactions are active only between fermions with different
spin polarization. In the homogeneous fluid this approximation on the
interactions leads to the 
RPA expression for the sound velocity \cite{paris-futur}. The results
of Baranov and Petrov \cite{baranov} are recovered by setting $g=0$ in
Eq.~(\ref{eul}). 

It is interesting to notice that Eq.~(\ref{eul}) for the velocity field
is irrotational, in analogy with the corresponding equation for a
Bose-condensed gas \cite{stringari_hydro_1996}. In this case it
contains explicitly the effect of  Fermi statistics in the
expression for the linearized local chemical potential.
Equations \refeq{cont} and \refeq{eul} apply only for
describing low-energy excitations at very low temperature, since
they do not include the coupling with thermal quasiparticle
excitations and the related damping. 

As an application of Eqs.~(\ref{cont}-\ref{eul}) we study  the
possibility of exciting the scissors mode in a superfluid Fermi gas.

\section{Scissors mode in a superfluid Fermi gas}
\label{due}
We consider an atomic cloud confined inside a slightly anisotropic
trap in the $xy$ plane, described by the confining potential 
\begin{equation}
V_{ext}(\vett r)=\frac{1}{2} m \omega_0^2 (1
+\varepsilon)x^2+\frac{1}{2} m \omega_0^2 (1 -\varepsilon)y^2
+\frac{1}{2} m \omega_z^2 z^2\;. 
\end{equation}
Within this geometry the scissors mode can be excited by a sudden
rotation of the trap through a small angle in the $xy$ plane. 
In the experiment by Marag{\`o} {\it et al.} \cite{onofrio} the anisotropy has been obtained by
adding  a small component in the $z$ direction to the 
magnetic field of the 
TOP trap and the scissors mode
has been excited by a sudden change of the sign of such field. 

For the study of the scissors mode of a superfluid Fermi gas  we
employ the technique of moments developed in \cite{francesca} (see
also \cite{david}).
 We define the
dynamical average of  a  variable $\chi(\vett r, \vett v)$ on
the total density profile $n(\vett r,t)$ as 
\begin{equation}
\langle \chi\rangle=\int d^3r\, \chi(\vett r, \vett
v)n(\vett r,t)\;. 
\end{equation}
For the scissors mode the relevant variable is $\chi=xy$, describing
the quadrupolar 
oscillation with $z$-component of the angular momentum $m_z=\pm 2$. By
making use of the equations of motion for the superfluid we obtain 
\begin{equation}
\partial_t \langle xy \rangle=\langle xv_y+yv_x\rangle
\label{sup-cont}
\end{equation}
from the continuity equation and 
\begin{equation}
\partial_t\langle xv_y+yv_x\rangle=-2\omega_0^2 \langle xy
\rangle
\label{sup-eul}
\end{equation}
from the linearized Euler equation.
A physical picture of the scissors mode can be obtained by observing
that the moment $\langle xy \rangle$ is related to the angle
$\theta$ of oscillation of the cloud
in the $xy$ plane, if  $\theta\ll 1$. For larger angles instead
the variable $\langle xy \rangle$ describes the usual quadrupolar
excitation of the cloud.

\begin{figure}
\centerline{\epsfig{file=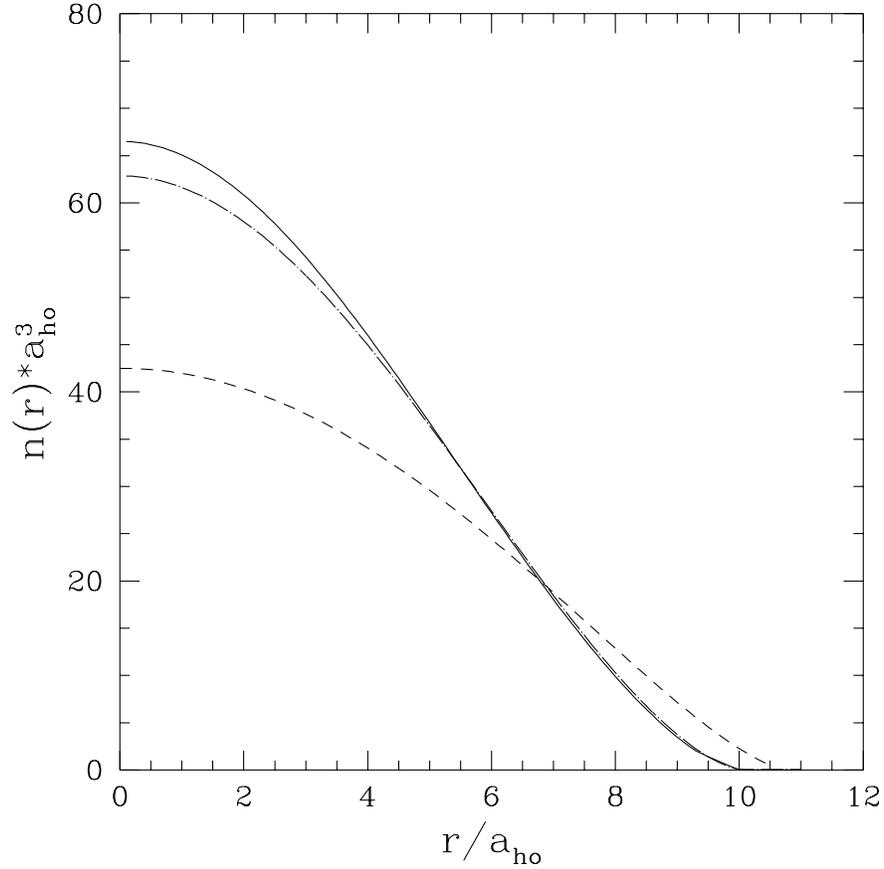,width=12cm}}
\caption{Equilibrium density profile at $T=0$ in spherical harmonic
confinement (with trapping frequency $\omega_{ho}=2 \pi \times 520$
s$^{-1}$) for a 
superfluid Fermi gas (solid line) and for a normal Fermi gas
(dot-dashed line) in the symmetric system at the same number of
particles $N=6.6 \times 10^4$. The two curves
are the result of  a local-density approximation using the chemical
potential of the homogeneous fluid, as evaluated in the first case
from the solution of the BCS equations, and in the second from
the non-interacting value  $\mu _{hom}[n]=\hbar^2 (3 \pi^2 n)^{2/3}/2m$.  
The equilibrium density profile of a non-interacting normal 
Fermi gas at the same $N$ is also shown (dashed line). }
\label{fig1}
\end{figure}

In deriving Eq.~\refeq{sup-eul}
we have made use of the expression for the equilibrium density profile in the
local-density (``Thomas-Fermi'') approximation
\cite{molmer,maddalena}, which for a 
symmetric two-component fluid  is the solution of $A [n_{eq}(\vett
r)/2]^{2/3}=\mu -V_{ext}(\vett r)+g n_{eq}(\vett r)/2$. We are here
exploiting the fact that even in the presence of weak
interactions the equilibrium density profile of the superfluid is well
approximated by the corresponding profile of a normal fluid at the
same number of particles. This property is  
illustrated in Figure~\ref{fig1} and is a consequence of the interplay
between  the
external confinement and interactions.
While interactions shift only slightly
the chemical potential of the homogeneous superfluid relative to that of the
normal fluid, in the presence of confinement they also enter 
in
the density profiles through the
position-dependent Hartree term. This second effect is dominant in
determining the profiles, as shown in Figure~\ref{fig1}.
As for confined Bose condensates, the fluid
is dilute but nevertheless interactions significantly modify the
density profile.

The solution of the coupled equations (\ref{sup-cont}) and (\ref{sup-eul})
yields for the frequency of
the mode the result
\begin{equation}
\omega^2=2 \omega_0^2\;.
\label{sup-sciss}
\end{equation}
 This value coincides with that for the ($n=0$ $l=2$) total-density surface
mode of a 
two-component normal Fermi gas in the collisional regime
\cite{ilaria2}. The same property  holds
for a Bose-condensed cloud \cite{david}.

In fact,
the scissors  mode is not affected by the
interactions: its frequency  coincides with that of the  ($n=0$ $l=2$)
  surface
mode in a single-component Fermi gas \cite{baranov,ilaria}. 
This is indeed a property of all  surface modes of total density
fluctuations \cite{ilaria2}. For these modes $\nabla \cdot 
\vett v=0$  and the equation of motion for the velocity field
reduces to 
\begin{equation}
m \partial_t^2\vett v(\vett r,t)=\nabla \left[\vett v(\vett r,t) \cdot 
\nabla [A 
(n_{eq}(\vett r)/2)^{2/3}+ g n_{eq}(\vett r)/2]\right]\;.
\end{equation}
By employing the Thomas-Fermi  form of the equilibrium  profile we have
\begin{equation}
m \partial_t^2\vett v(\vett r,t)=-\nabla \left[\vett v(\vett r,t) \cdot \nabla
V_{ext}(\vett r)\right]\;. \label{incomp}
\end{equation}
This equation shows  that only the external confinement
determines the frequency of these modes. Equation~(\ref{incomp}) is
independent of the statistics: it describes as well the surface modes of
a classical gas in the hydrodynamic regime and of a Bose-Einstein
condensate at $T=0$ \cite{griff_ss}.

\section{Scissors modes for a normal Fermi gas}
\label{tre}
We shall now contrast the result \refeq{sup-sciss} for the superfluid
with the excitation frequencies of the scissors modes in a 
two-component Fermi fluid in the normal state. 
At a temperature higher than the superfluid transition temperature
$T_{sup}$ the gas is described by the Vlasov-Landau equation for
the Wigner distribution functions $f_{\vett p}^\sigma(\vett r,t)$ of
each of the components:
\begin{equation}
\left(\partial_t +\frac{\vett p}{m}\cdot \nabla_{\vett r}-
\nabla_{\vett r}(V_{ext}(\vett r) + g n_{\bar \sigma}(\vett r,t))\cdot 
\nabla_{\vett p}\right)f^\sigma_{\vett p}(\vett r,t)= I_{coll} [f_{\vett
p}^\sigma(\vett r,t)]\;. 
\end{equation} 
The density profile of each component is defined in terms of
 the Wigner function as $n_\sigma(\vett r,t)=\int d^3p f^\sigma_{\vett
p}(\vett r,t)/(2 \pi 
\hbar)^3$ and $I_{coll}$ is the collision
integral.

The average of the variable $\chi(\vett r,\vett v)$ on each
 component $\sigma$ of the gas is defined as 
\begin{equation}
\langle \chi\rangle_\sigma=\frac{1}{(2 \pi \hbar)^3}\int d^3r\int d^3p
\, \chi(\vett r, \vett 
v)f^{\sigma}_{\vett p}(\vett r,t)\;. 
\end{equation} 
The equation of motion for $ \langle xy \rangle_\sigma$ is now coupled
to both the longitudinal and the transverse excitations and to the kinetic
tensor fluctuations, yielding the following set of equations:
\begin{eqnarray}
\partial_t \langle xy \rangle_\sigma&=&\langle
xv_y+yv_x\rangle_\sigma\;,
\label{norm-cont} \\
\partial_t\langle xv_y-yv_x\rangle_\sigma&=&-2\varepsilon\omega_0^2(C
\langle xy 
\rangle_\sigma+(1-C)\langle xy \rangle_{\bar \sigma})\;,\\
\partial_t\langle xv_y+yv_x\rangle_\sigma&=&2\langle \Pi_{xy}\rangle_\sigma/m-2\omega_0^2(C \langle xy
\rangle_\sigma+(1-C)\langle xy \rangle_{\bar \sigma})
\end{eqnarray}
and
\begin{equation}
\partial_t\langle \Pi_{xy}\rangle_\sigma=-mC\omega_0^2\left[\langle
xv_y+yv_x\rangle_\sigma +\varepsilon\langle xv_y-yv_x\rangle_\sigma
\right]-\langle \Pi_{xy}\rangle_\sigma/\tau \;.
\label{norm-kin}
\end{equation}
We have defined the current density $n_\sigma \vett v_\sigma=\int d^3p\,\vett p
f^\sigma_{\vett p}(\vett r,t)/m (2 \pi
\hbar)^3$ and  the $xy$ component of the kinetic stress tensor
$\Pi_{xy}^\sigma=\int d^3p\,p_x p_y f^\sigma_{\vett 
p}(\vett r,t)/m (2 \pi 
\hbar)^3$, and 
 made use  of the Thomas-Fermi
Ansatz
$An_{eq}^\sigma(\vett r)^{2/3}=C(E_F-V_{ext}(\vett r))$
for the equilibrium density profile in a symmetric system, as
already employed in \cite{ilaria2}. The constant $C=(E_F^0/E_F)^2$,
where $E_F$ is the true Fermi energy and $E_F^0$ is that of the
non-interacting Fermi gas, measures the strength of the interactions
in its deviations from unity. The approximations that we have made
require 
 $|C-1|\ll 1$ and  $T\ll T_F$; the latter condition is well compatible with the condition $T>T_{sup}$ in a weakly
interacting fluid. 
Notice
that the motions of the two components of the gas
are coupled for $C\ne 1$ .

Collisions have been included in
Eq.~\refeq{norm-kin} within a single-relaxation-time approximation
by setting
\begin{equation}
\langle\Pi_{xy}I_{coll} \rangle_\sigma=-\langle\Pi_{xy} \rangle_\sigma/\tau\;.
\end{equation}
This approximation
subsumes damping by scattering against impurities and by inter-species
scattering (see \cite{ilaria2} for a full discussion).

In the collisionless regime ($\omega\tau \gg 1$), equations
(\ref{norm-cont}-\ref{norm-kin}) can be combined to yield the
following coupled differential equations for $\langle
xy\rangle_\sigma$:
\begin{equation}
(\partial_t^4+4 C \omega_0^2 \partial_t^2 +4 \varepsilon^2\omega_0^4C^2)\langle
xy\rangle_\sigma +2 \omega_0^2(1-C)(- \partial_t^2 +2
\varepsilon^2\omega_0^2)\langle xy\rangle_{\bar \sigma}=0\;.
\end{equation}
Solution by diagonalization yields two scissors modes 
associated with 
total density 
fluctuations and two further modes associated with   concentration
fluctuations, at frequencies given by
\begin{equation}
\omega^2=2 \omega_0^2 \left\{\begin{array}{l} 2C\mp (1-C) \\
\varepsilon^2 C(C\pm (1-C))/(2C\mp (1-C))\end{array}\right.
\end{equation}
The appearance of soft modes with a frequency proportional to
$\varepsilon$ is a 
peculiarity of a non-superfluid system. 

In the collisional regime ($\omega\tau \ll 1$),
Eqs.~(\ref{norm-cont}-\ref{norm-kin}) instead yield
\begin{equation}
(\partial_t^3+2C\omega_0^2\partial_t)\langle
xy\rangle_\sigma +2(1-C) \omega_0^2 \partial_t\langle xy\rangle_{\bar \sigma}=0\;.
\end{equation}
This has  two solutions for  non-zero frequencies of the scissor modes at
\begin{equation}
\omega^2=2\omega_0^2(C\pm (1-C))\;.
\end{equation}
These agree with  the ($n=0$ $l=2$) surface modes already obtained by a
different approach in \cite{ilaria2}. As expected, the frequency of
the total-density mode reproduces the result in Eq.~(\ref{sup-sciss}).

\section{Summary and concluding remarks}
\label{quattro}	
 In this work we have extended the equations of motion for  density
 fluctuations in an
 inhomogeneous Fermi gas in the superfluid state at $T=0$ to include 
 interactions at a mean-field (RPA) level within a local-density
 approximation. In the appropriate limits our approach reproduces
 Anderson result \cite{anderson_rpa,paris-futur} for the speed of the
 Bogolubov-Anderson phonon in the homogeneous superfluid and those by
 Baranov and Petrov \cite{baranov} for a confined non-interacting superfluid.

 As an application, we have computed the frequency of
 the scissors mode for a fermionic superfluid gas.
 By comparing the result with those for a normal Fermi gas in the
 collisionless regime, we have shown that 
 the measurement of  the scissors modes could be used
 as a signal for the superfluid transition, since well below the
 transition temperature the soft transverse modes are suppressed. 
 However, these modes are already non-propagating in a normal Fermi
 fluid in the collisional regime.

 We have also found that the excitation frequency of the scissors
 mode in a superfluid Fermi gas coincides with that of a normal Fermi
 gas in the collisional regime. This result
 is understood in the fermionic case as related to the
 opening of the gap 
 which stabilizes the first-sound mode. As is the case for all 
 surface modes, interactions do not shift the frequency of the
 scissors mode in the superfluid.

 While we have worked here within a local-density approximation,
 further work should be addressed to treat in a fully quantal way the
 dynamical properties of the inhomogeneous fluid of present interest. 

\acknowledgements
One of us (AM) wishes to
acknowledge useful discussions with Professor A.~Griffin and with
Dr. D.~Gu\`ery-Odelin.


\begin{thebibliography}{10}

\bibitem{fermi_degen}
B. DeMarco and D.~S. Jin, Science {\bf 285},  1703  (1999).

\bibitem{gabriele}
M.~O. Mewes, G. Ferrari, F. Schreck, A. Sinatra, and C. Salomon, Phys. Rev. A
  {\bf 61},  011403(R)  (2000).

\bibitem{stoof_prl}
H.~T.~C. Stoof, M. Houbiers, C.~A. Sackett, and R.~G. Hulet, Phys. Rev. Lett.
  {\bf 76},  10  (1996).

\bibitem{stoof_marianne}
M. Houbiers, R. Ferweda, H.~T.~C. Stoof, W.~I. McAlexander, C.~A. Sackett, and
  R.~G. Hulet, Phys. Rev. A {\bf 56},  4864  (1997).

\bibitem{baranov_bcs}
M.~A. Baranov, Y. Kagan, and M.~Y. Kagan, JETP Lett. {\bf 64},  301  (1996).

\bibitem{david}
D. Gu{\`e}ry-Odelin and S. Stringari, Phys. Rev. Lett. {\bf 83},  4452  (1999).

\bibitem{onofrio}
O.~M. Marag{\`o}, S.~A. Hopkins, J. Arlt, E. Hodby, G. Hechenblaikner, and C.~J.
  Foot, Phys. Rev. Lett. {\bf 84},  2056  (2000).

\bibitem{forster}
D. Forster, {\em Hydrodynamic Fluctuations, Broken Symmetry, and Correlation
  Functions} (Benjamin, Reading, 1975).

\bibitem{baranov}
M.~A. Baranov and D.~S. Petrov, cond-mat/9901108.

\bibitem{ilaria}
M. Amoruso, I. Meccoli, A. Minguzzi, and M.~P. Tosi, Eur. Phys. J. D {\bf 7},
  441  (1999).

\bibitem{anderson_rpa}
P.~W. Anderson, Phys. Rev. {\bf 112},  1900  (1958).

\bibitem{bogolubov_sound}
N.~N. Bogolubov, V.~V. Tolmachev, and D.~V. Shirkov, {\em New Method in the
  Theory of Superconductivity} (Academy of the Sciences of the U.S.S.R.,
  Moscow, 1958).

\bibitem{nambu_rpa}
Y. Nambu, Phys. Rev. {\bf 117},  648  (1960).

\bibitem{griffin_rpa}
R. C{\^o}t{\'e} and A. Griffin, Phys. Rev. B {\bf 48},  10404  (1993).

\bibitem{leggett}
A.~J. Leggett, Phys. Rev. {\bf 140},  1869  (1965).

\bibitem{migdal}
A.~I. Larkin and A.~B. Migdal, Sov. Phys. JETP {\bf 17},  1146  (1963).


\bibitem{paris-futur}
A. Minguzzi, G. Ferrari, and Y. Castin, in preparation.

\bibitem{leggett2}
A.~J. Leggett, Phys. Rev. {\bf 147},  119  (1966).


\bibitem{martin_rpa}
P.~C. Martin,  in {\em Superconductivity},  edited by R.~D. Parks (Dekker, New
  York, 1969), Vol.~1,  p.\ 371.



\bibitem{ilaria2}
M. Amoruso, I. Meccoli, A. Minguzzi, and M.~P. Tosi, Eur. Phys. J. D {\bf 8},
  361  (2000). By a symmetric vapour we mean equal number and equal
confinement for each of the two components: this is the best situation
to obtain pairing \cite{stoof_marianne}. 

\bibitem{stringari_hydro_1996}
S. Stringari, Phys. Rev. Lett. {\bf 77},  2360  (1996).

\bibitem{francesca}
D. Gu{\`e}ry-Odelin, F. Zambelli, J. Dalibard, and S. Stringari, Phys. Rev. A
  {\bf 60},  4851  (1999).

\bibitem{molmer}
K. M{\o}lmer, Phys. Rev. Lett. {\bf 80},  1804  (1998).

\bibitem{maddalena}
M. Amoruso, A. Minguzzi, S. Stringari, M.~P. Tosi, and L.Vichi, Eur. Phys. J. D
  {\bf 4},  261  (1998).

\bibitem{griff_ss}
A. Griffin, W.-C. Wu, and S. Stringari, Phys. Rev. Lett. {\bf 78},  1838
  (1997).

\end{thebibliography}

\end{document}